\documentclass[final,5p,times,twocolumn]{elsarticle}

\journal{Physics Letters B}
\biboptions{sort&compress}

\usepackage{xcolor}

\definecolor{redd}{rgb}{0.8, 0.1,0.2}
\definecolor{navy}{rgb}{0.05, 0.23,0.75}
\usepackage{bbm}
\usepackage{amsfonts}
\usepackage{amsmath,amssymb}
\usepackage{mathrsfs}
\usepackage{graphicx}
\usepackage{url}
\usepackage[colorlinks=true,citecolor=navy,linkcolor=redd,urlcolor=navy,
            anchorcolor=blue]{hyperref}
\hypersetup{colorlinks=true}

\newcommand{\nn}{\nonumber}

\newcommand{\be}{\begin{equation}}
\newcommand{\ee}{\end{equation}}
\newcommand{\bea}{\begin{eqnarray}}
\newcommand{\eea}{\end{eqnarray}}
\newcommand{\bc}{\begin{center}}
\newcommand{\ec}{\end{center}}

\begin{document}

\begin{frontmatter}
		
\title{
Phenomenological implications of a class of non-invertible selection rules
}

\author[sissa,infn]{Motoo Suzuki}
\ead{msuzuki@sissa.it}
	
\author[ictp]{Ling-Xiao Xu}
\ead{lxu@ictp.it}

\affiliation[sissa]{organization={SISSA International School for Advanced Studies},
                    addressline={Via Bonomea 265, 34136, Trieste, Italy}}
\affiliation[infn]{organization={INFN, Sezione di Trieste},
                   addressline={Via Valerio 2, 34127, Italy}}
\affiliation[ictp]{organization={Abdus Salam International Centre for Theoretical Physics},
                   addressline={Strada Costiera 11, 34151, Trieste, Italy}}

\begin{abstract}
We demonstrate that non-invertible fusion algebras give rise to a class of selection rules with genuine organizing power in particle-physics models, which we call non-invertible selection rules (NISRs). We identify the algebraic structures that distinguish NISRs from ordinary group-based selection rules and from their generic explicit breaking.
As a minimal example, we study a singlet-scalar extension of the Standard Model and show that the NISR based on Fibonacci fusion rules leads to distinctive scattering patterns. We further apply the Ising fusion rules to a supersymmetric flipped SU(5) model, addressing the doublet–triplet splitting problem. In this construction, the NISR plays a crucial role in forbidding the $\mu$-term, thereby evading a no-go result for ordinary Abelian symmetries. We anticipate that this class of selection rules can be applied to a wide range of models beyond the Standard Model. To facilitate such applications, we provide a pedagogical review of radiative violation of NISRs in the appendices, with a dark matter model also being discussed. 
\end{abstract}
	
\end{frontmatter}

\section{Introduction}
The notion of global symmetries has been generalized significantly since the seminal work~\cite{Gaiotto:2014kfa}. A particularly striking generalization is the so-called non-invertible symmetries, whose symmetry defects obey fusion rules rather than group multiplication; see e.g.~\cite{Cordova:2022ruw, Shao:2023gho, Schafer-Nameki:2023jdn, Costa:2024wks, Brennan:2023mmt} for reviews. While such structures have been familiar in two-dimensional rational conformal field theories~\cite{Chang:2018iay, Thorngren:2019iar, Thorngren:2021yso}, their generalizations to four-dimensional quantum field theories~\cite{Choi:2021kmx, Kaidi:2021xfk} and applications in particle physics~\cite{Choi:2022jqy, Cordova:2022ieu, Putrov:2023jqi, Choi:2023pdp, Cordova:2022fhg, Cordova:2023her, Cordova:2024ypu, Delgado:2024pcv, Kobayashi:2024yqq, Kobayashi:2024cvp, Funakoshi:2024uvy, Kobayashi:2025znw, Cao:2024lwg} have only been uncovered recently. Celebrated examples include the non-invertible symmetry associated with the Adler-Bell-Jackiw anomaly in quantum electrodynamics~\cite{Choi:2022jqy, Cordova:2022ieu} and the higher-form generalizations in axion-Maxwell theory~\cite{Choi:2022fgx, Yokokura:2022alv, Hidaka:2024kfx, DelZotto:2024ngj}.

A central implication for particle physics is that selection rules governing particle interactions need not obey group laws. In this paper, we study a class of selection rules induced by non-invertible fusion algebras that are exact at tree level in perturbation theory~\cite{Kaidi:2024wio} (see also~\cite{Heckman:2024obe}), which we call non-invertible selection rules (NISRs). We aim to show their usefulness in uncovering some peculiar features in particle-physics models, which would otherwise be difficult to explain using the ordinary group-based selection rules. Remarkably, such structures already arise in simple weakly-coupled extensions of the Standard Model (SM), which are the focus of this work.

Through concrete models, we demonstrate three layers of structure encoded by NISRs.
\begin{enumerate}
\item \emph{Beyond ordinary group-based selection rules.} Because the fusion product of two elements may contain several channels, the set of tree-level allowed operators is selected in a way that does not follow from ordinary group-based charge conservation.
\item \emph{Structured symmetry breaking.} Although not itself a group-based symmetry, a NISR can be organized through an auxiliary lifted group~\cite{Suzuki:2025bxg, Suzuki:2025kxz}, in which non-invertibility is traded for a restricted set of explicit-breaking terms~\cite{Xu:2026nwh}. This restricted set contains more information than generic symmetry breaking: it preserves tree-level exactness~\cite{Kaidi:2024wio}, the closure property that at tree level, gluing two NISR-allowed amplitudes yields another NISR-allowed amplitude. This consistency condition implies that not every pattern of symmetry-breaking operators can arise from NISRs.
\item \emph{Controlled radiative violation.} When particles labeled by non-invertible elements propagate in loops, they can generate vertices that are absent at tree level. This loop-induced violation is not arbitrary: the algebraic structure of the fusion rules determines, through the loop-induced groupification procedure~\cite{Kaidi:2024wio}, the perturbative loop order at which a given vertex can first be generated.
\end{enumerate}
Therefore, a NISR is not merely a statement of symmetry breaking, but a finer algebraic structure that selects tree-level seed operators, constrains their composition, and controls their radiative violation.

\section{Framework}
Let $A=\{\mathbbm{1}, x, y, z, \cdots\}$ be a finite set of basis elements obeying the commutative fusion rules 
\be
x\otimes y=y\otimes x=\sum_{z\in A} N^z_{x,y} z\ , 
\label{eq:fusionrules}
\ee
where $\mathbbm{1}$ is the identity element and $N^z_{x,y}\in \mathbb{Z}_{\geq 0}$ are the structure constants. The basis elements need not be invertible; in particular, a fusion product may contain several channels, so it cannot be interpreted as group multiplication.

Following Ref.~\cite{Kaidi:2024wio}, each dynamical field $\phi_i$ is labeled by an element $x_i\in A$, and the conjugate field $\bar{\phi}_i$ carries the dual label $\bar x_i$ such that the identity appears in the fusion product of $x_i$ and $\bar x_i$, i.e., $\mathbbm{1}\prec x_i\otimes\bar x_i$.
Here $\prec$ means that the corresponding element appears with nonzero multiplicity in the fusion product.
Kinetic terms are thus always allowed for all dynamical fields. More generally, a local interaction $\prod_i\phi_i$ is allowed at tree level if and only if $\mathbbm{1}\prec\prod_i x_i$. This rule extends to all tree-level amplitudes by the closure property of tree-level gluing. 

At higher loop orders, this tree-level rule is modified by the loop-induced groupification procedure~\cite{Kaidi:2024wio}, which we review in the appendices. Radiative corrections can generate some tree-level forbidden vertices, while others remain forbidden. Hence, the NISR is progressively relaxed in radiative corrections rather than arbitrarily broken. At first sight, it may seem counterintuitive that tree-level exact NISRs can be violated radiatively.  This is clarified by the spurion analysis of Refs.~\cite{Suzuki:2025bxg, Suzuki:2025kxz}, whose underlying intuition has recently been developed into a systematic algorithm~\cite{Xu:2026nwh}. 

In phenomenological applications, NISRs are often implemented as the absence of certain tree-level couplings~\cite{Kobayashi:2024cvp}, reducing the number of free parameters in the theory. Since Lagrangian couplings are not physical observables, such zeros are defined only after choosing an operator basis and renormalization scheme; changing schemes redefines couplings and counterterms, while leaving the relations among physical observables unchanged. 
Following Coleman and Weinberg~\cite{Coleman:1973jx}, one may define these zeros by imposing
renormalization conditions on the effective potential $V_{\rm eff}$, or more generally on the one-particle-irreducible (1PI) effective action.  For a vertex involving fields $\phi_{i_1},\ldots,\phi_{i_n}$ with fusion labels $x_{i_1},\ldots,x_{i_n}$, we impose
\be
\left.
\frac{\partial^n V_{\rm eff}}
{\partial\phi_{i_1}\cdots\partial\phi_{i_n}}
\right|_{\phi=\phi_R}
=0
\qquad\text{when}\qquad
\mathbbm{1}\not\prec x_{i_1}\otimes\cdots\otimes x_{i_n},
\label{eq:RC}
\ee
where $\phi_R$ is the chosen renormalization point, typically the origin but possibly a nonzero value when needed. Eq.~\eqref{eq:RC} defines the renormalized NISR-forbidden coupling to vanish at $\phi_R$; it does not imply that the corresponding derivative of $V_{\rm eff}$ vanishes away from this point, nor that loop-induced effects are absent with generic kinematics in physical processes. These conditions fix the counterterms in the chosen scheme. At lowest order, they reduce to the tree-level absence of NISR-forbidden couplings.

\section{A real scalar and Fibonacci fusion rules}

We first illustrate NISRs in the SM extended by a gauge-singlet real scalar $s$, whose phenomenology has been widely studied~\cite{OConnell:2006rsp, Chen:2014ask, Buttazzo:2015bka, He:2016sqr, Cao:2017oez, Corbett:2017ieo, Li:2019tfd, Haisch:2020ahr, Adhikari:2020vqo, Dawson:2021jcl, Lewis:2024yvj}. We consider the classical Lagrangian
\be
\mathcal{L}=\mathcal{L}_{\text{SM}}+\frac{1}{2} (\partial_\mu s)^2-\frac{m_s^2}{2} s^2-\frac{\lambda_3}{3!} s^3-\frac{\lambda_4}{4!} s^4-\frac{\lambda_{sh}}{2} s^2 |H|^2 \;,
\label{eq:model_s}
\ee
where we assume $m_s^2, \lambda_4, \lambda_{sh}>0$ for simplicity.

From the viewpoint of the ordinary $\mathbb{Z}_2$ transformation $s\to -s$, this interaction pattern at tree level is unconventional. An exact $\mathbb{Z}_2$ symmetry would forbid all the terms with odd powers of $s$, while the presence of $s^3$ explicitly breaks it. Therefore, ordinary $\mathbb{Z}_2$ reasoning does not explain why other $\mathbb{Z}_2$-breaking terms, such as $\lambda_1 s$ and $\lambda^\prime_{sh} s |H|^2$, are absent at tree level. Without an additional selection rule, this absence is merely a tuning in the usual 't Hooft sense~\cite{tHooft:1979rat}. Relatedly, the missing single-$s$ vertices are nevertheless generated radiatively. For instance, the 1PI vertex associated with $s|H|^2$ first appears at one loop by gluing the tree-level $s^3$ and $s^2|H|^2$ vertices, 
\be
\Gamma_{s|H|^2}^{\rm 1-loop} \sim \frac{\lambda_3^{\rm tree}\lambda_{sh}^{\rm tree}}{16\pi^2}.
\label{eq:coup_fib}
\ee
This raises the question whether a selection rule can allow the couplings $m_s^2,\lambda_3,\lambda_4,\lambda_{sh}$ at tree level while forbidding $\lambda_1, \lambda^\prime_{sh}$ in the classical Lagrangian, thereby making the physical process $s\to H^\dagger H$ naturally loop suppressed compared to $s\to s s$ and $s s \to H^\dagger H$.~\footnote{Notice that some particles have to be off-shell due to kinematics. However, our focus here is on the dynamics, namely the interactions between particles.}
This loop order is not explained by the $\mathbb Z_2$ group, i.e., it contains no algebraic structure that distinguishes \(s^3\) as a tree-level seed from \(s|H|^2\) as a radiative descendant. From the $\mathbb Z_2$ viewpoint, the reverse order would be equally consistent.

We propose to use the class of NISRs introduced in~\cite{Kaidi:2024wio} to address the above question. For the real-scalar model, we use the Fibonacci fusion algebra~\footnote{Conventionally, the Fibonacci fusion rules are used to describe anyons in condensed matter physics~\cite{trebst2008short}. Recently, the same algebra also appeared in flavor models explaining the Yukawa texture zeros based on $\mathbb{Z}_2$ gauging of a $\mathbb{Z}_3$ symmetry~\cite{Kobayashi:2024cvp, Funakoshi:2024uvy, Kobayashi:2025znw}.}, whose fusion rules are given by
\be
\mathbbm{1}\otimes \mathbbm{1}=\mathbbm{1}, \quad\quad \mathbbm{1}\otimes \tau=\tau\otimes \mathbbm{1}=\tau, \quad\quad \tau\otimes \tau= \mathbbm{1} \oplus \tau\;.
\label{eq:fib}
\ee
The element $\tau$ is non-invertible, as $\tau\otimes\tau$ contains more than one fusion channel.
In the model, we label the real scalar $s$ by the element $\tau$ and all SM particles by $\mathbbm{1}$. It is easy to check that all the terms in Eq.~\eqref{eq:model_s} are consistent with Fibonacci fusion rules, i.e., the identity element is produced in the fusion product of the corresponding particles in each of the interactions. By contrast, operators linear in $s$ are forbidden. Therefore, the Fibonacci fusion rules realize precisely the desired interaction pattern as in Eq.~\eqref{eq:model_s}, from which Eq.~\eqref{eq:coup_fib} follows as a consequence.

\emph{Selection rules beyond $\mathbb{Z}_2$.} 
The successive self-fusions of $\tau$ give rise to the Fibonacci sequence $\{0,1,1,2,3,5,8,\ldots\}$, where the $n$-th number counts the multiplicity of $\mathbbm{1}$ obtained in the fusion product of $\tau^n$~\cite{trebst2008short}. This gives a selection rule distinct from a $\mathbb Z_2$ group.  While a $\mathbb Z_2$ symmetry distinguishes odd from even powers of $s$, the Fibonacci fusion algebra distinguishes single-$s$ operators $s\mathcal O_{\rm SM}$ from multi-$s$ operators $s^n\mathcal O_{\rm SM} \ (n>1)$, where only the latter are allowed.
Here $\mathcal O_{\rm SM}$ denotes an operator built only from SM fields. Since all SM fields are labeled by $\mathbbm{1}$, the internal structure of $\mathcal O_{\rm SM}$ is not constrained.

From another perspective, it is useful to view the Fibonacci-allowed operators $s^n\mathcal O_{\rm SM}$ with $n>1$ as $s^2\left(s^{n-2}\mathcal O_{\rm SM}\right)$. Here $s^2$ is a composite operator labeled by $\tau^2$, which is decomposed as $1\oplus \tau$. This multi-channel structure makes the full operator Fibonacci-allowed even when the remaining factor is not allowed independently.  For instance, when $n=3$, both $s^3\mathcal O_{\rm SM}$ and $s^2$ are allowed, whereas the remaining factor $s\mathcal O_{\rm SM}$ is not.
More generally, let two fields $\phi_1$ and $\phi_2$ carry distinct fusion labels, while a composite operator $\mathcal O$ has several fusion channels. Then both $\phi_1^\dagger\mathcal O$ and $\phi_2^\dagger\mathcal O$ may contain the identity in their fusion products, without requiring $\phi_1^\dagger\phi_2$ to contain the identity. This pattern cannot be
realized by an ordinary Abelian charge assignment---if both $\phi_1^\dagger\mathcal O$ and $\phi_2^\dagger\mathcal O$ are charge-neutral, then $\phi_1^\dagger\phi_2$ is also charge-neutral. The multi-channel fusion structure is precisely what allows NISRs to select tree-level operator patterns beyond ordinary Abelian charge conservation.

\emph{Fibonacci as structured $\mathbb{Z}_2$ breaking.}
In general, the set of operators allowed by NISRs obeys the closure property known as tree-level exactness~\cite{Kaidi:2024wio}. An arbitrary set of explicit symmetry-breaking operators need not satisfy this condition. 
For instance, let us consider $s |H|^2$ and $s^3 |H|^2$ in the real-scalar model. If $s |H|^2$ is allowed by a NISR, one can construct a tree-level process whose external legs match $s^3 |H|^2$, thereby implying that $s^3 |H|^2$ must also be allowed. Therefore, a pattern in which $s|H|^2$ is allowed while $s^3|H|^2$ is forbidden is \emph{incompatible} with tree-level exactness, and cannot be realized by \emph{any} NISRs. 
By contrast, from the viewpoint of a broken $\mathbb Z_2$ symmetry alone, all explicit $\mathbb Z_2$-breaking patterns are on the same footing, since this closure structure is not encoded.
Thus, from the $\mathbb Z_2$-breaking perspective, the Fibonacci NISR realizes a structured rather than generic breaking pattern.

\emph{Controlled radiative corrections.} 
As we will analyze in the appendices, the procedure of loop-induced groupification~\cite{Kaidi:2024wio} suggests that the Fibonacci fusion algebra reduces to the trivial group at one-loop order, thereby allowing the tree-level forbidden process $s\to H^\dagger H$ to be generated at one-loop order. Here, the commutator structure of the Fibonacci fusion algebra determines $s^3$ as a tree-level seed and $s|H|^2$ as a radiative descendant at one-loop order.

\begin{figure}[t]
\centering
\includegraphics[width=\linewidth]{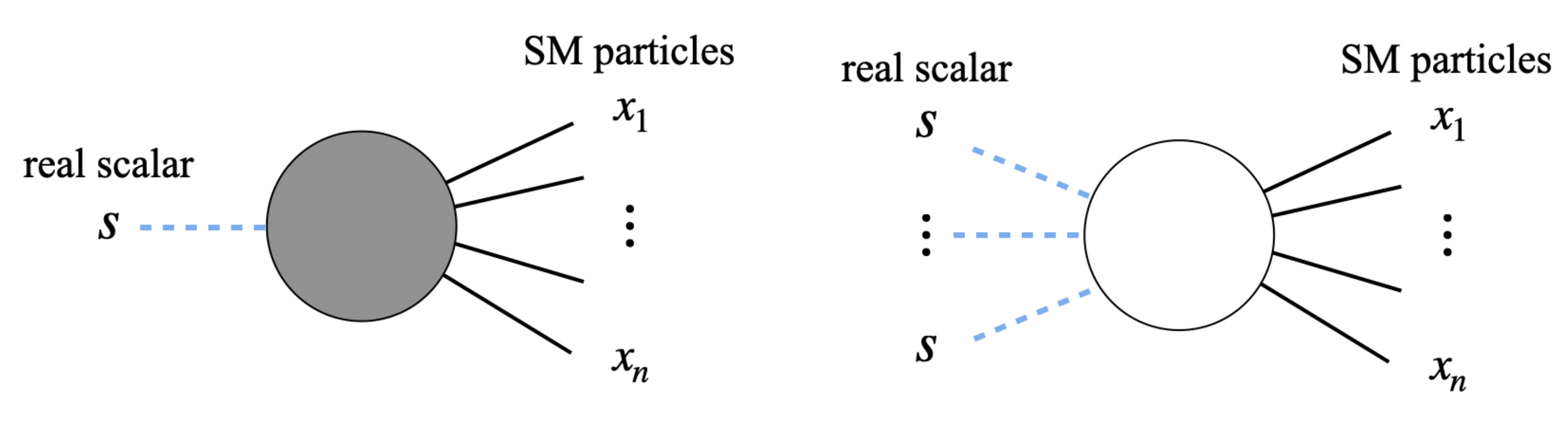}
\caption{A cartoon showing the scattering processes of the real scalar $s$ and the SM particles whose interactions are constrained by the Fibonacci fusion rules as in Eq.~\eqref{eq:fib}. All the single-$s$ scatterings must be loop induced (denoted by the gray blob), while the multi-$s$ scatterings are present at either tree or loop levels (denoted by the white blob) in perturbation theory. If the real scalar is discovered by future experiments, one can potentially test the Fibonacci fusion rules in these scattering processes. }
\label{fig:scattering_S}
\end{figure}

Overall, the central physical prediction is that \emph{all} the single-$s$ scattering processes must be loop-induced, while the multi-$s$ scatterings can be present at tree-level; see in Fig.~\ref{fig:scattering_S}. This yields testable correlations among physical cross sections, which are controlled by the few parameters in the classical Lagrangian, Eq.~\eqref{eq:model_s}, with counterterms fixed by the renormalization conditions in Eq.~\eqref{eq:RC}. Once the number of independently measured processes exceeds the number of free parameters, the setup can be tested through these nontrivial correlations among observables. A quantitative study is pursued elsewhere.

To conclude, our point is that a NISR carries more algebraic data than a broken group. This additional structure constrains which breaking pattern is selected in the first place, i.e., which operators serve as tree-level seeds, why the selection rule is preserved exactly in tree-level processes, and at which lowest loop order the forbidden vertices first appear. These features are not fixed by the broken $\mathbb Z_2$ symmetry alone, since different breaking operators carry the same $\mathbb Z_2$ spurion charge. The conventional broken-$\mathbb Z_2$ spurion analysis can still organize some radiative effects once the breaking terms are introduced by hand, but it does not itself contain the algebraic data needed to determine the breaking pattern with the same level of detail as a NISR.

\section{The $\mu$ term and Ising fusion rules}

We further illustrate the power of NISRs beyond ordinary group-based selection rules by solving the doublet--triplet splitting problem in the supersymmetric flipped $SU(5)\times U(1)_\chi$ model~\cite{Barr:1981qv, Derendinger:1983aj, Antoniadis:1987dx, Barr:1988yj}. In this setup, the missing-partner mechanism~\cite{Masiero:1982fe, Grinstein:1982um} keeps the electroweak Higgs doublets light while lifting their colored-triplet partners to the grand-unification (GUT) scale. However, once the Majorana neutrino mass term is incorporated, no ordinary invertible symmetry can forbid the bilinear $\mu$-term 
\be
W_\mu = \mu\, h\bar h, 
\ee
so it is generically allowed, reintroducing the doublet--triplet splitting problem~\cite{Kuflik:2010dg}. We show that the Ising fusion rules evade this no-go result, where the non-invertibility relaxes the usual Abelian charge-conservation conditions that always allow $h\bar h$.

\emph{Setup.} In the supersymmetric flipped $SU(5)\times U(1)_\chi$ model, the three matter generations $F_i,\bar f_i,\bar e_i$ $(i=1,2,3)$, the GUT-breaking Higgs fields $H,\bar H$, and the electroweak Higgs fields $h,\bar h$ carry the
quantum numbers
\begin{equation}
\begin{array}{c|ccc|cc|cc}
& F_i & \bar f_i & \bar e_i & H & \bar H & h & \bar h \\ \hline
SU(5)
& \mathbf{10}
& \bar{\mathbf 5}
& \mathbf 1
& \mathbf{10}
& \overline{\mathbf{10}}
& \mathbf 5
& \bar{\mathbf 5}
\\
U(1)_\chi
& +1 & -3 & +5 & +1 & -1 & -2 & +2
\end{array}
\label{eq:fields}
\end{equation}
with the flipped embedding $F_i\supset(Q_i,\bar d_i,N_i)$, $\bar f_i\supset(\bar u_i,L_i)$, and $\bar e_i$ the $SU(2)_L$-singlet charged lepton. The electroweak Higgs doublets are embedded as $h_d\subset h$ and $h_u\subset\bar h$, together with
colored-triplet partners. The GUT symmetry is broken by vacuum expectation values (VEV) along the SM-singlet directions with
$\langle H\rangle=\langle\bar H\rangle\sim M_{32}$, which Higgs $SU(5)\times U(1)_\chi$ to $SU(3)_c\times SU(2)_L\times U(1)_Y$. The required superpotential is
\be
W=\underbrace{FFh+F\bar f\bar h+\bar f\bar e h}_{W_Y}+\underbrace{HHh+\bar H\bar H\bar h}_{W_{DT}}+\underbrace{\Lambda^{-1}F\bar H F\bar H}_{W_N}.
\label{eq:W}
\ee
The terms in $W_Y$ generate the SM Yukawa couplings---$FFh$ gives the down-quark masses, $F\bar f\bar h$ gives the up-quark and Dirac-neutrino masses, including $F\bar f\bar h\supset Q\bar u\,h_u+NLh_u$, and $\bar f\bar e h$ gives
the charged-lepton masses.  The missing-partner couplings in $W_{DT}$ give GUT-scale masses to the colored-triplet components of $h,\bar h$ through their mixing with the triplet components of $H,\bar H$, while the electroweak doublets remain unpaired.  Finally, after $\bar H$ acquires its VEV, the dimension-five operator $W_N$ generates Majorana masses for
the right-handed neutrinos, $M_R\sim M_{32}^2/\Lambda$; together with the Dirac mass $m_D=y_u v_u$ from $F\bar f\bar h$, this gives the type-I seesaw relation $m_\nu\sim m_D^2/M_R$~\cite{Minkowski:1977sc,Yanagida:1979as, Glashow:1979nm, Gell-Mann:1979vob}.

\emph{No-go and its evasion.} The obstruction of~\cite{Kuflik:2010dg} follows directly from charge conservation.  For
an ordinary non-$R$ $U(1)$ symmetry, the operators $FFh$, $\bar H\bar H\bar h$, and $F\bar H F\bar H$ in Eq.~\eqref{eq:W} gives $q_h=-2q_F, q_{\bar h}=-2q_{\bar H}, q_F+q_{\bar H}=0$, and hence $q_h+q_{\bar h}=0$. Thus the bare bilinear $h\bar h$ is neutral and allowed; the same conclusion holds modulo $n$ for ordinary $\mathbb Z_n$ symmetries. For a $U(1)_R$ symmetry, every superpotential term must carry $R(W)=2$. The terms $FFh$ and $HHh$ imply $R(F)=R(H)$, while $F\bar H F\bar H$ gives $R(F)+R(\bar H)=1$, hence $R(H)+R(\bar H)=1$.
Together with the conditions $R(h)=2-2R(H)$ and $R(\bar h)=2-2R(\bar H)$ implied by $W_{DT}$, one obtains $R(h)+R(\bar h)=2$, so the superpotential bilinear $h\bar h$ is again allowed.

One must also forbid the following dangerous higher-dimensional operators
\begin{equation}
\mathcal O_m
=
\Lambda^{-(2m-1)}(H\bar H)^m h\bar h,
\qquad m\ge1 ,
\label{eq:tower}
\end{equation}
which after GUT breaking induce
\[
\mu_{\rm ind}^{(m)}
\sim
\frac{\langle H\rangle^{2m}}{\Lambda^{2m-1}}
=
\delta^{2m-1}M_{32},
\qquad
\delta\equiv \frac{M_{32}}{\Lambda}.
\]
For typical $\delta\gtrsim10^{-2}$, these contributions are far above the TeV scale. Notice that a $U(1)_R$ symmetry removes this tower. Indeed, the relations above yield
\begin{align}
R\!\left[(H\bar H)^m h\bar h\right]
&=
m\bigl(R(H)+R(\bar H)\bigr)+R(h)+R(\bar h)
\nonumber\\
&=m+2 .
\label{eq:Rtower}
\end{align}
Hence all operators $\mathcal O_m$ with $m\ge1$ are forbidden by $U(1)_R$, while $h\bar h$ is still allowed. This is the only operator that \emph{must} be removed by the NISR.

We use the Ising fusion algebra to forbid the $\mu$-term $W_\mu$. The algebra is defined for a set of three basis elements $\{\mathbbm{1}, \epsilon, \sigma\}$ with the fusion rules
\begin{align}
\epsilon\otimes \epsilon=\mathbbm{1}, \quad \epsilon\otimes \sigma=\sigma\otimes \epsilon=\sigma, \quad \sigma\otimes \sigma=\mathbbm{1}\oplus\epsilon\ ,
\label{eq:Ising1}
\end{align}
where $\mathbbm{1}\otimes x=x\otimes \mathbbm{1}=x$ is implicitly understood for $x=\{\mathbbm{1},\epsilon,\sigma\}$. 
The element $\sigma$ is non-invertible because $\sigma\otimes\sigma$ contains multiple fusion channels; consequently, there exists no element $x$ such that $\sigma\otimes x=x\otimes \sigma= \mathbbm{1}$. 
In the model, we label the fields
\begin{align}
F,h &\sim \mathbbm 1, & H,\bar H &\sim \sigma, & \bar f,\bar e,\bar h &\sim \epsilon .
\label{eq:IsingAssign}
\end{align}
An operator is allowed if the fusion product of its labels contains $\mathbbm 1$. It is straightforward to verify that all the terms in Eq.~\eqref{eq:W} are allowed. By contrast, the bare $\mu$-term $W_\mu$ is forbidden.  
Again, let us focus on the three allowed operators $FFh$, $\bar H\bar H\bar h$ and $F\bar H F\bar H$ that were used to derive the no-go obstruction for non-$R$ symmetries, it is useful to rewrite them as
\be
(FF)h, \quad (\bar H\bar H) \bar h, \quad (\bar H \bar H) (FF)\ .
\ee
The key point is that Eq.~\eqref{eq:Ising1} does not impose ordinary additive charge conservation among these operators. Since $\bar H \bar H$ contains two channels $\mathbbm{1}$ and $\epsilon$ in the fusion product, the labeling $FF, h\sim \mathbbm 1$, $\bar h\sim \epsilon$ can realize the desired pattern.

It is important that the NISR is not used alone to forbid all the dangerous terms. Indeed, if the two terms in $W_{DT}$ are allowed by non-$R$ NISR, $\mathcal O_2=\Lambda^{-3}(HHh)(\bar H\bar H\bar h)$ is also allowed.
Thus the evasion is genuinely complementary: the $U(1)_R$ symmetry forbids the induced operators $\mathcal O_m$ with $m\ge1$, while the Ising NISR forbids the remaining bare bilinear term $h\bar h$. The combined $U(1)_R\otimes{\rm Ising}$ selection rule therefore removes the full dangerous $\mu$-term tower while preserving the Yukawa, missing-partner, and neutrino-mass operators.

\emph{Proton decay and R-parity violation.} The same Ising NISR also forbids the dimension-five operators that can induce proton decay $FFF\bar f, F\bar f\bar f\bar e$ and the leading R-parity-violating operators $H\bar f\bar h, HFF\bar f, H\bar f\bar f\bar e$. With Eq.~\eqref{eq:IsingAssign}, all these operators are forbidden. By contrast, these operators are difficult to forbid with ordinary invertible symmetries~\cite{Kuflik:2010dg}. This illustrates an additional advantage of the Ising NISR.

\section{Discussion and outlook}
With two simple models, we have shown that NISRs provide nontrivial organizing principles for particle-physics models. 
Unlike generic symmetry breaking, where it is not obvious whether any meaningful selection rule remains, NISRs retain algebraic information that continues to organize the operator spectrum.
This point is particularly clear in our second example, where interpreting the Ising NISRs merely as ``symmetry breaking'' would not be regarded as a resolution to the no-go result.

Here we have formulated NISRs at the Lagrangian level, which is useful but involves choices of fields and operator basis. A more invariant formulation can instead be given directly in terms of scattering amplitudes. In this perspective, the external dynamical particles are labeled by elements of a fusion algebra, and the NISR selects the allowed contact amplitudes as tree-level seeds, from which all other tree- and loop-level amplitudes are generated inductively by gluing. In this formulation, the Lagrangian is only a bookkeeping device that manifests NISRs. Such a viewpoint is advocated e.g. in~\cite{Kaidi:2024wio, Suzuki:2025bxg, Suzuki:2025kxz}.

We see several directions to extend our results.
\begin{itemize}
\item It is known that NISRs can arise as low-energy imprints of string theory. Although this perspective was partially appreciated in the early seminal works~\cite{Hamidi:1986vh, Font:1988nc, Kobayashi:1995py}, it has been made more explicit and systematic only recently~\cite{Kaidi:2024wio, Heckman:2024obe}. From a bottom-up perspective, one may instead study how theories constructed from distinct NISRs are related along renormalization-group flows, on which NISRs can impose nontrivial constraints. Some preliminary aspects were sketched in~\cite{Suzuki:2025kxz}.

\item Radiative aspects of NISRs are systematically captured by ``groupification''~\cite{Kaidi:2024wio}, under the natural assumptions that the fusion algebra is faithfully realized and captures all particle quantum numbers. This, however, is not the typical case for realistic models, where particles do carry additional quantum numbers. The interplay between different quantum numbers can obstruct groupification, potentially leading to stronger loop-order suppression~\cite{Suzuki:2025bxg, Suzuki:2025kxz}. This remains largely unexplored. 
\end{itemize}

Our results may hint at a paradigm shift, and we hope that they will stimulate further exploration into the realm of non-invertible particle physics.

\section*{Acknowledgment}
We would like to thank Hao Y. Zhang for bringing~\cite{Kaidi:2024wio} to our attention and for collaboration on the related work on spurion analysis~\cite{Suzuki:2025bxg}, Di Zhang for helpful comments on the texture zeros and flavor models, Rajath Radhakrishnan for helpful clarifications on fusion rings and categories. 
M.S. is supported by the MUR project 2017L5W2PT.
M.S. also acknowledges the European Union - NextGenerationEU, in the framework of the PRIN Project “Charting unexplored avenues in Dark Matter” (20224JR28W).
The work of L.X.X. is partially supported by European Research Council (ERC) grant n.101039756. L.X.X. would like to thank ITP-CAS for their warm hospitality at the early stage of this work. 

During the preparation of this work, the authors used OpenAI's ChatGPT to improve the language and readability of the manuscript. After using this tool, the authors reviewed and edited the content as needed and take full responsibility for the content of the publication.

\appendix
\renewcommand{\theHfigure}{appendix.\Alph{section}.\arabic{figure}}
\renewcommand{\theHtable}{appendix.\Alph{section}.\arabic{table}}
\renewcommand{\theHequation}{appendix.\Alph{section}.\arabic{equation}}

\begin{center}
   \textbf{\large APPENDICES \\[.2cm] ``Phenomenological implications of a class of non-invertible selection rules'' }\\[.2cm]
\end{center}

In these appendices, we first discuss a dark matter (DM) model based on the Ising fusion rules, demonstrating the broader applicability of our method. 

Next, we provide a pedagogical review of loop-induced groupification~\cite{Kaidi:2024wio} and explain how it captures the radiative breaking of NISRs. Along the way, we present an intuitive way of visualizing it, which facilitates practical applications to particle phenomenology and further motivates the connection with spurion analysis~\cite{Suzuki:2025bxg, Suzuki:2025kxz, Xu:2026nwh}. 

Finally, we present a concrete example in which groupification identifies the exact unbroken symmetry, while obtaining the same result by direct inspection of the Lagrangian would be impractical.

\noindent
\section{Dark matter and Ising fusion rules}
\label{app:DMIsing}
\setcounter{equation}{0}
\setcounter{figure}{0}
\setcounter{table}{0}

Let us introduce a gauge-singlet real scalar $s$ and a gauge-singlet Majorana fermion $\chi$, whose interactions are described by the following classical Lagrangian: 
\bea
    \mathcal{L}&=\mathcal{L}_{\text{SM}}
    +\frac{1}{2}(\partial_\mu s)^2-\frac{m_s^2}{2}s^2-\frac{\lambda_{s}}{4!}s^4-\frac{1}{2}\lambda_{s h}s^2|H|^2\nn\\
    &+i\chi^\dagger \bar\sigma^\mu\partial_\mu\chi -\left(\frac{m_\chi}{2} \chi^2
    +\frac{y_{s\chi}}{2}s\, \chi^2+
    \text{h.c.}\right)\;,
    ~\label{eq:model_2}
\eea 
where $m_s^2, \lambda_s, \lambda_{sh}>0$.
In particular, we assume that the terms $s$, $s^3$, and $s|H|^2$ are not allowed in the classical Lagrangian. It follows that the scattering processes $s\to s s$ and $s\to H^\dagger H$ must be loop-induced. For instance, $s\to s s$ can be induced at the one-loop level by a triangle loop of $\chi$ due to the tree amplitude $s\to \chi\chi$. In turn, $s\to H^\dagger H$ can be induced at the two-loop level by gluing together the legs of $s$ in the one-loop amplitude $s\to s s$ and the tree amplitude $s s \to H^\dagger H$. 

Let us analyze the ``symmetry" structure of the classical Lagrangian in Eq.~\eqref{eq:model_2}. In the paradigm of ordinary symmetries, it is difficult to explain the following two features simultaneously.
\begin{enumerate}
\item The interactions of $s$ with the SM particles respect a $\mathbb{Z}_2$ symmetry where $s\to -s$, but those with $\chi$ do not. For instance, due to the presence of the $s\chi^2$ term in Eq.~\eqref{eq:model_2}, it is not possible to assign a $\mathbb{Z}_2$-odd charge to the scalar $s$.
Since there is no $\mathbb{Z}_2$ symmetry protecting the interactions of $s$, the absence of the terms $s, s^3, s|H|^2$ in Eq.~\eqref{eq:model_2} is not justified. 
\item However, Eq.~\eqref{eq:model_2} enjoys an invertible $\mathbb{Z}_2$ symmetry (i.e., fermion parity) under which $\chi\to -\chi$ while all the other particles are charged even. Hence $\chi$ is a natural DM candidate.
This $\mathbb{Z}_2$ symmetry is exact at all loop orders in perturbation theory.
\end{enumerate}

Here we seek a paradigm shift: we point out that the above two features can be \emph{coherently} understood in a single framework using the non-invertible Ising fusion rules, 
\begin{equation}
\begin{array}{c|@{\hspace{0.8cm}}c@{\hspace{0.8cm}}c@{\hspace{0.8cm}}c}
\otimes & \mathbbm{1} & \epsilon & \sigma \\
\hline
\mathbbm{1} & \mathbbm{1} & \epsilon & \sigma \\
\epsilon   & \epsilon   & \mathbbm{1} & \sigma \\
\sigma     & \sigma     & \sigma & \mathbbm{1}\oplus\epsilon
\end{array}
\label{eq:Ising_appendix}
\end{equation}
as we already introduced in our second example.
We label the fields in Eq.~\eqref{eq:model_2} as follows:
\begin{equation}
s \sim \epsilon,\qquad \chi \sim \sigma,\qquad \Phi_{\rm SM} \sim \mathbbm{1},
\end{equation}
where $\Phi_{\rm SM}$ denotes all SM fields.
Here are the reasons why the Ising fusion rules are useful in understanding Eq.~\eqref{eq:model_2}.
\begin{enumerate}
\item It is straightforward to check that all the terms in Eq.~\eqref{eq:model_2} are consistent with the Ising fusion rules, hence they are tree-level exact. Furthermore, all the terms (up to dimension four) that are consistent with the Ising fusion rules are already present in the classical Lagrangian of Eq.~\eqref{eq:model_2}. 
This explains the peculiar feature that only $s\chi^2$ is present in the classical Lagrangian, while $s, s^3, s|H|^2$ are not. Naively, $s\chi^2$ would be forbidden by the $\mathbb{Z}_2$ symmetry where $s\to -s$. However, this term is allowed once the fusion rule of the non-invertible element $\sigma$ is taken into account. 
\item The all-loop exact $\mathbb{Z}_2$ symmetry (fermion parity), which ensures DM stability, arises from ``groupification'' of the non-invertible Ising fusion rules~\cite{Kaidi:2024wio}. In particular, groupification states that at sufficiently high loop orders, NISRs reduce to a finite Abelian group that remains exact up to all loop orders. We will review ``groupification'' later.
This all-loop exact $\mathbb{Z}_2$ symmetry should not be confused with the Ising $\mathbb{Z}_2$ associated with the element $\epsilon$, where $\epsilon\otimes \epsilon=\mathbbm{1}$. The latter is already violated at one loop through radiative corrections, as signaled by the generation of the process $s\to s s$.
\end{enumerate}

\begin{figure}[t]
\centering
\includegraphics[width=\linewidth]{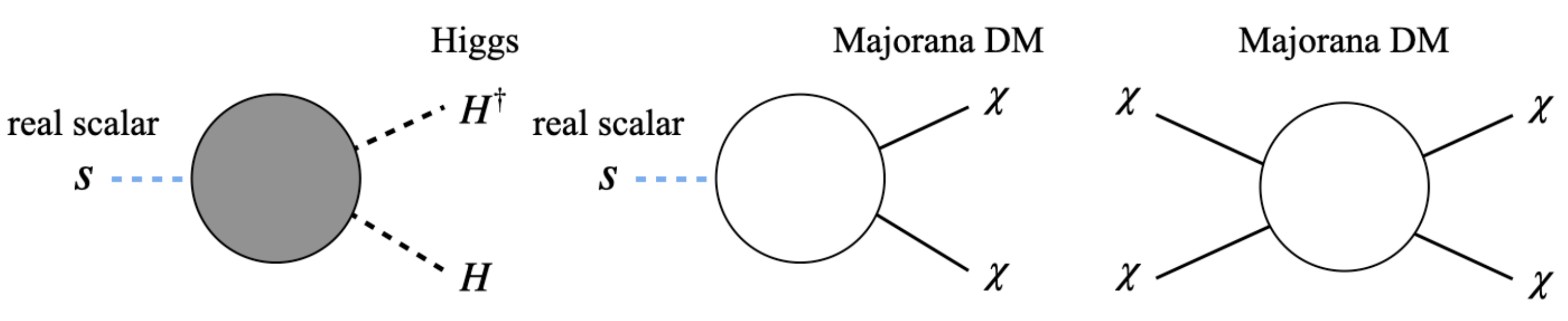}
\caption{Schematic scattering processes in the DM model constrained by the Ising fusion rules. The process $s\to H^\dagger H$ is absent at tree level and must be loop induced, as denoted by the gray blob. By contrast, $s\to \chi\chi$ and $\chi\chi\to\chi\chi$ can be present at tree level, as denoted by the white blobs. These processes are relevant for DM annihilation and direct detection phenomenology.}
\label{fig:scattering_model2}
\end{figure}

The Ising fusion rules lead to a characteristic pattern for the interactions relevant to DM phenomenology. As illustrated schematically in Fig.~\ref{fig:scattering_model2}, the processes $s\to \chi\chi$ and $\chi\chi\to\chi\chi$ can arise at tree level, whereas $s\to H^\dagger H$ is absent at tree level and can only be generated radiatively. Phenomenologically, when kinematically allowed, DM annihilation through $\chi\chi\to ss$ can proceed at tree level, while direct detection is mediated by loop-induced $s$--$H$ mixing. This results in a characteristic pattern in the annihilation and direct-detection rates, which is not present in generic models with or without a conventional $\mathbb Z_2$ symmetry acting on $s$.
It would therefore be interesting to systematically explore the parameter space of this ``non-invertible Ising DM'' scenario as a benchmark, following the analyses in~\cite{Krnjaic:2015mbs, Matsumoto:2018acr, Chen:2024njd, Krnjaic:2017tio}.

\noindent
\section{A lightning review on loop-induced groupification of NISRs}
\label{app:groupification}
\setcounter{equation}{0}
\setcounter{figure}{0}
\setcounter{table}{0}

It was recently pointed out that NISRs, when applied in perturbation theory, are increasingly violated as the loop order increases and eventually reduce to those of a finite Abelian group at sufficiently high loop orders. The entire procedure is called the \emph{groupification} of the original NISRs.

The radiative violation of NISRs, and hence the process of groupification, occurs because some basis elements become \emph{indistinguishable} at loop orders. More specifically, let $x$ and $y$ be two different basis elements of the set $A$, each labeling at least one particle in the theory.
We say that $x$ and $y$ cannot be distinguished at a given loop order $L$ if there exists some element $\omega$ constructed from the products of the conjugate pairs such that:
\begin{align}
  &  x\prec \omega \otimes y\ , \label{eq:groupification_1}\\
  & \text{where}\quad \omega \prec \bigotimes_{i=1}^L
  z_i,~z_i\prec u\otimes\bar{u} \quad\text{for some $u\in A$}\;.\label{eq:groupification_2}
\end{align}
Here, $\prec$ means that the fusion product on the right-hand side contains the element on the left-hand side. At sufficiently high loop orders (i.e., taking the limit $L\to\infty$), the set made from the elements $w$ stabilizes, and accordingly, the fusion algebra reduces to a finite Abelian group~\cite{Kaidi:2024wio}. 

More systematically, this construction can be expressed in terms of the ``commutator'' set of the fusion algebra. We define
\begin{align}
\mathrm{Com}(A)
:=
\left\{
z\in A \,\middle|\, z\prec u\otimes\bar u
\ \text{for some } u\in A
\right\},
\label{eq:ComA_def}
\end{align}
namely the set of basis elements that can appear from fusing a label with its
conjugate. Its $L$-fold product is
\begin{align}
\mathrm{Com}(A)^L
:=
\left\{
\omega\in A \,\middle|\,
\omega\prec z_1\otimes\cdots\otimes z_L,\quad
z_i\in\mathrm{Com}(A)
\right\}.
\label{eq:ComA_L_def}
\end{align}
Thus $\mathrm{Com}(A)^L$ collects the labels that can be generated by $L$ internal conjugate pairs running in loops.
This notation provides a practical loop-order criterion for an $n$-point vertex $\phi_1\phi_2\cdots\phi_n$, where the fields carry labels $x_1,x_2,\ldots,x_n\in A$. Decompose the fusion product
\begin{equation}
X:=x_1\otimes x_2\otimes\cdots\otimes x_n
\end{equation}
into the basis elements of $A$. It follows that:
\begin{enumerate}
\item if $\mathbbm{1}\prec X$, the vertex is allowed at tree level;
\item if $\mathbbm{1}\not\prec X$, but $X$ contains an element of $\mathrm{Com}(A)$, the vertex can first be generated at one loop;
\item more generally, if $X$ contains no element of $\mathrm{Com}(A)^k$ for $k<L$, but contains an element of
$\mathrm{Com}(A)^L$, the vertex can first be generated at $L$-loop order;
\item if $X$ contains no element of the stabilized set $\mathrm{Com}(A)^\infty$, the vertex is forbidden by the all-loop exact groupified symmetry.
\end{enumerate}

\begin{figure}[t]
\centering
\includegraphics[width=\linewidth]{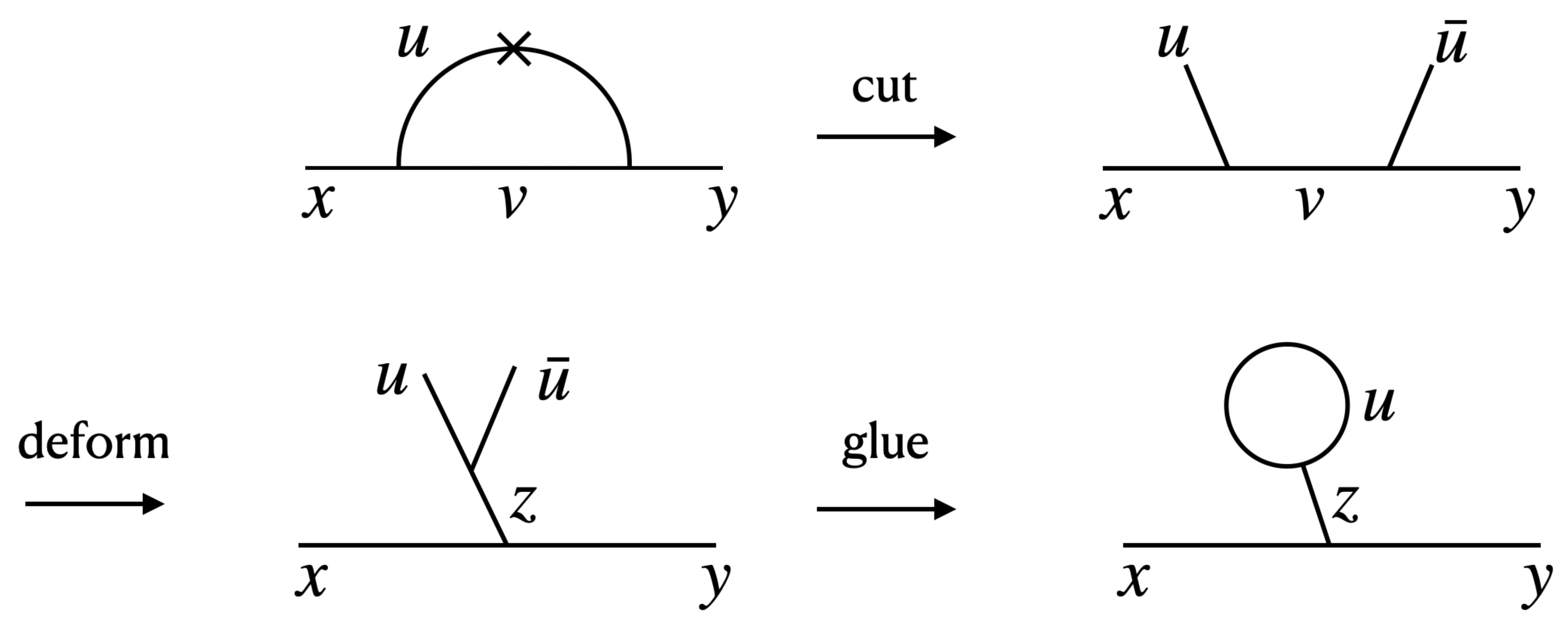}
\caption{Diagrammatic illustration of Eqs.~\eqref{eq:groupification_1} and~\eqref{eq:groupification_2}, where the two particles labeled by $x$ and $y$ become indistinguishable at the one-loop level due to quantum corrections. For this to happen, it is important that $u$ is non-invertible; otherwise, the element $z$ can only be identity, which cannot be used to identify $x$ and $y$.}
\label{fig:groupification}
\end{figure}

Eqs.~\eqref{eq:groupification_1} and~\eqref{eq:groupification_2} can also be understood diagrammatically. For instance, as shown in Fig.~\ref{fig:groupification}, the particles labeled by $x$ and $y$ are indistinguishable at the one-loop order --- namely $L=1$ --- if there exists an element $z\prec u\otimes \bar{u}$ such that $y\prec z\otimes x$. (However, notice that $x$ and $y$ are distinct at tree level, where the two-point vertex mixing $x$ and $y$ is forbidden.) In a sequence, the diagrams in Fig.~\ref{fig:groupification} are as follows:
\begin{enumerate}
\item Due to quantum correction, the particle labeled by $x$ can split into two virtual particles labeled by $u$ and $v$, respectively. Here, the elements satisfy the relation $v\prec x\otimes u$ in the fusion algebra.
\item We cut open the internal line for the virtual particle labeled by $u$. The original one-loop diagram becomes a tree-level one, where the elements from the left to right satisfy the relations $v\prec x\otimes u, \ y\prec v\otimes \bar{u}$ in the fusion algebra. From these relations, it is easy to see that $y\prec x \otimes u\otimes \bar{u}$. 
\item We deform the tree diagram obtained in the last step, such that the order of fusions is changed, i.e., $z\prec u\otimes \bar{u}, \ y\prec x \otimes z$. Notice that $z$ does not have to correspond to a dynamical particle in the theory, but it may be related to the label $v$ through the associators (i.e., the $F$-symbols). A systematic understanding on the implications of the $F$-symbol in particle phenomenology is left for future work.
\item Finally, we glue together the conjugate pair $u$ and $\bar{u}$. We see that the elements $x$ and $y$ become indistinguishable by dressing the element $z$, which in turn is produced by $u\otimes \bar{u}$ at the one-loop order. 
\end{enumerate}
The same reasoning applies to higher loop orders, where one can always cut open the internal lines, such that the loop diagram reduces to a tree-level one, whose external legs contain the original external lines plus some number of additional conjugate pairs. Here, the number of additional conjugate pairs matches the internal lines being cut. After suitable deformation, one can glue the conjugate pairs together, and then one can explicitly see that some elements become indistinguishable precisely because of the dressing of conjugate pairs, i.e., Eqs.~\eqref{eq:groupification_1} and~\eqref{eq:groupification_2}. 

From the perspective of the spurion analysis, one can absorb the elements $z_i$ produced by the conjugate pairs $u\otimes \bar{u}$, which label dynamical particles, into the definition of coupling constants. (Therefore, one can choose to replace all conjugate pairs with the identity element in the spurion analysis.) For related discussions of NISRs from near-group fusion algebras and beyond, see~\cite{Suzuki:2025bxg, Suzuki:2025kxz, Xu:2026nwh}.

\textbf{Fibonacci case.} For the NISRs from the Fibonacci fusion algebra, the process of groupification is completed already at the one-loop order. Namely, the set of elements $w$ produced by the conjugate pairs stabilizes at the one-loop order, i.e., $\{w\}=\{\mathbbm{1},\tau\}$ at $L=1$ where both elements are obtained from the fusion product $\tau\otimes\tau$; see Eq.~\eqref{eq:groupification_2}. By dressing $\tau\in \{w\}$, we find $\mathbbm{1}\prec \tau\otimes \tau$ as in Eq.~\eqref{eq:groupification_1}, i.e., the two elements $\mathbbm{1}$ and $\tau$ become indistinguishable and equivalent at the one-loop order. As a result, the Fibonacci fusion algebra reduces to the trivial group, which consists only of the identity as its group element. 

\begin{figure}[t]
\centering
\includegraphics[width=\linewidth]{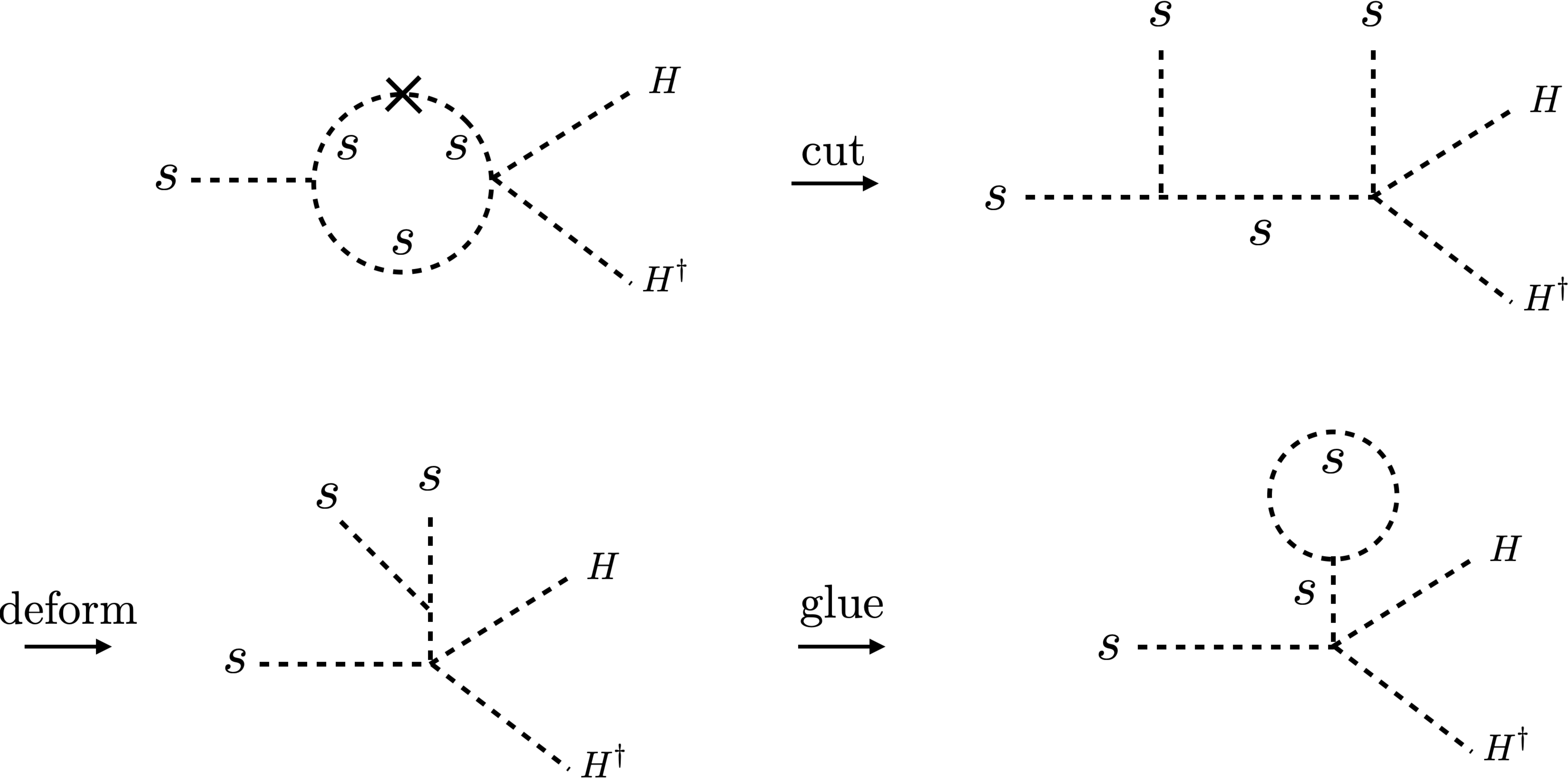}
\caption{
Diagrammatic illustration of groupification for the Fibonacci fusion algebra, which is concretely realized in the real scalar model. Specifically, here we consider the three-point vertex $s|H|^2$.
}
\label{fig:fibo2}
\end{figure}

In the real gauge-singlet scalar model, the argument of groupification can be visualized by the diagrams shown in Fig.~\ref{fig:fibo2}. Specifically, we consider the three-point vertex $s|H|^2$, where the real scalar $s$ is labeled by the element $\tau$ while the Higgs doublet $H$ (hence $H^\dagger H$) is labeled by $\mathbbm{1}$. Clearly, such a vertex is forbidden at tree level, but it is generated at the one-loop order. Correspondingly, the relevant process $s\to |H|^2$ can occur precisely because $\tau$ and $\mathbbm{1}$ become indistinguishable and equivalent.
In a sequence, the diagrams in Fig.~\ref{fig:fibo2} are as follows:
\begin{enumerate}
\item Due to quantum correction, the real scalar $s$ can split into two virtual scalars $s$. Here, the corresponding elements satisfy the relation $\tau\prec \tau\otimes \tau$ in the Fibonacci fusion algebra.
\item We cut open the internal line for $s$. The original one-loop diagram becomes a tree-level one, where the elements from left to right satisfy the relations $\tau\prec \tau\otimes \tau, \ \mathbbm{1}\prec \tau\otimes \tau$ in the Fibonacci fusion algebra. From these relations, it is easy to see that $\mathbbm{1}\prec \tau \otimes \tau\otimes \tau$. Notice that $\tau$ is self-conjugate, namely $\bar{\tau}=\tau$.
\item We deform the tree diagram obtained in the last step, such that the order of fusions is changed.
\item Finally, we glue together the conjugate pair back to restore the loop order. We see that the elements $\tau$ and $\mathbbm{1}$ become indistinguishable by dressing the element $\tau$, which in turn is produced by $\tau\otimes \tau$ at the one-loop order. 
\end{enumerate}

From the viewpoint of the spurion analysis, we can choose to absorb the element $\tau$ in the last step --- which is produced by the fusion of a conjugate pair $\tau\otimes \tau$ --- into the definition of the coupling constant for the three-point vertex $s |H|^2$, i.e., $\lambda^\prime_{sh}\sim \tau$. In turn, it is related to the tree couplings $\lambda_3$ and $\lambda_{sh}$ respectively for the vertices $s^3$ and $s^2 |H|^2$. Following~\cite{Suzuki:2025bxg}, we label the tree couplings as $\lambda_3\sim \tau$ and $\lambda_{sh}\sim \mathbbm{1}$, such that the relation between the couplings $\lambda^\prime_{sh}\sim \lambda_3 \lambda_{sh}/(16 \pi^2)$ is consistent with the fusion rule $\tau\prec \tau\otimes \tau$.

\textbf{Ising case.} For the NISRs from the Ising fusion algebra, the groupification is again completed at the one-loop order, since the set of elements $w$ produced by the conjugate pairs stabilizes at the one-loop order, i.e., $\{w\}=\{\mathbbm{1},\epsilon\}$ at $L=1$, where $\mathbbm{1}\prec\epsilon \otimes \epsilon$, $\mathbbm{1}\prec \sigma\otimes \sigma$, and $\epsilon\prec \sigma\otimes \sigma$. These equations are in the same form as Eq.~\eqref{eq:groupification_2}. Furthermore, by dressing with $\epsilon$, we find that $\mathbbm{1}\prec \epsilon \otimes \epsilon$ and $\sigma\prec \sigma\otimes \epsilon$, which implies that at the one-loop order $\epsilon$ becomes indistinguishable and equivalent to $\mathbbm{1}$, while $\sigma$ remains distinct; see Eq.~\eqref{eq:groupification_1}. Consequently, at the one-loop order and higher, the Ising fusion algebra reduces to a $\mathbb{Z}_2$ group, where the $\mathbb{Z}_2$-odd charge is generated by $\sigma$.  

\begin{figure}[t]
\centering
\includegraphics[width=\linewidth]{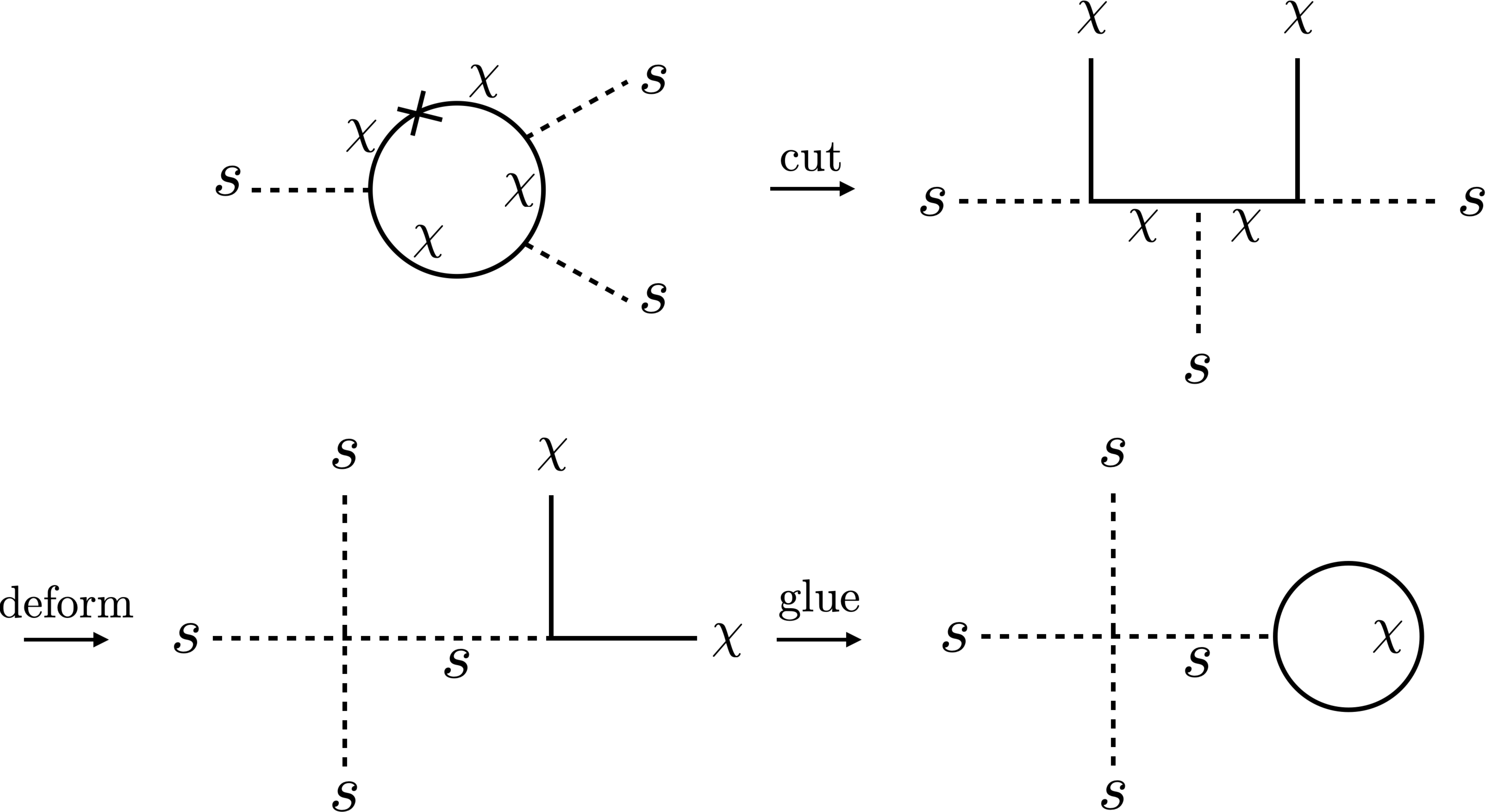}
\caption{
Diagrammatic illustration of groupification for the Ising fusion algebra, which is concretely realized in the DM model. Specifically, we consider the three-point vertex $s^3$ here.
}
\label{fig:ising2}
\end{figure}

In the DM model, the argument of groupification can be visualized by the diagrams shown in Fig.~\ref{fig:ising2}. Specifically, we consider the three-point vertex $s^3$ that is forbidden at tree level but is generated at the one-loop order. In the Ising fusion algebra, the mediator $s$ and the DM $\chi$ are labeled by the elements $\epsilon$ and $\sigma$, respectively. Hence, the process $s\to s s$ can occur precisely because $\epsilon$ and $\mathbbm{1}$ become indistinguishable and equivalent at the one-loop order.

From the viewpoint of the spurion analysis, we can choose to absorb the element $\epsilon$ in the last step --- which is produced by the fusion of the conjugate pair $\sigma\otimes \sigma$ as in Fig.~\ref{fig:ising2} --- into the definition of the coupling constant for the three-point vertex $s^3$, denoted by $\lambda_{3s}\sim \epsilon$. In turn, it is related to the tree coupling $y_{s\chi}$ for the vertex $s\chi^2$. Following~\cite{Suzuki:2025bxg}, we label the tree couplings as $y_{s\chi}\sim \epsilon$, such that the relation between the couplings $\lambda_{3s}\sim y^3_{s\chi}/(16 \pi^2)$ is consistent with the fusion rule $\epsilon\prec \epsilon\otimes \epsilon\otimes \epsilon$.

\noindent
\section{Groupification and cosine $\mathbb{Z}_M/\mathbb{Z}_2$ fusion algebras}
\label{app:groupification-cosine}
\setcounter{equation}{0}
\setcounter{figure}{0}
\setcounter{table}{0}

For simple fusion algebras such as Fibonacci or Ising, the exact groupified symmetry can often be identified directly from the corresponding Lagrangian. However, this becomes increasingly impractical for more general fusion algebras, where the interaction structure can be complicated, and the unbroken symmetry may not be immediately manifest. The groupification algorithm instead provides a systematic way to determine the all-loop exact symmetry directly from the fusion data itself.

We illustrate this point using the $\mathbb{Z}_M/\mathbb{Z}_2$ fusion algebra, which can be regarded as the discrete avatar of the cosine symmetry~\cite{Shao:2023gho}. Spurion analysis for this class of fusion algebras with generic $M$ was studied in~\cite{Suzuki:2025kxz}. Applying the groupification algorithm, it is easy to find that for even $M$ the all-order exact symmetry is $\mathbb{Z}_2$, whereas for odd $M$ all elements are identified and the groupified symmetry is trivial. 
As a case study, let us focus on $M=10$, where the fusion rules are
\begin{equation}
\begingroup
\setlength{\arraycolsep}{8pt}
\resizebox{\columnwidth}{!}{$
\begin{array}{c|cccccc}
\otimes & [g^0] & [g^1] & [g^2] & [g^3] & [g^4] & [g^5] \\
\hline
[g^0] & [g^0] & [g^1] & [g^2] & [g^3] & [g^4] & [g^5] \\

[g^1] & [g^1] & [g^0]\oplus[g^2] & [g^1]\oplus[g^3] & [g^2]\oplus[g^4] & [g^3]\oplus[g^5] & [g^4] \\

[g^2] & [g^2] & [g^1]\oplus[g^3] & [g^0]\oplus[g^4] & [g^1]\oplus[g^5] & [g^2]\oplus[g^4] & [g^3] \\

[g^3] & [g^3] & [g^2]\oplus[g^4] & [g^1]\oplus[g^5] & [g^0]\oplus[g^4] & [g^1]\oplus[g^3] & [g^2] \\

[g^4] & [g^4] & [g^3]\oplus[g^5] & [g^2]\oplus[g^4] & [g^1]\oplus[g^3] & [g^0]\oplus[g^2] & [g^1] \\

[g^5] & [g^5] & [g^4] & [g^3] & [g^2] & [g^1] & [g^0]
\end{array}
$}
\endgroup
\label{eq:Z10_Z2_fusion_table}
\end{equation}
Apart from the identity $[g^0]$ and $[g^{5}]$, all elements are non-invertible; all elements are self-conjugate since $[g^0]\prec[g^j]\otimes[g^j]$. Notice that the generic fusion rules take the form $[g^i]\otimes [g^j]=[g^{i+j}]\oplus [g^{i-j}]$, identical to the trigonometric identity of cosine functions, i.e., $[g^j]\sim \cos(2\pi j/M)$.

As one can read off directly from the fusion rules~\eqref{eq:Z10_Z2_fusion_table}, 
\begin{equation}
\mathrm{Com}\left(\mathbb{Z}_{10}/\mathbb{Z}_{2}\right)=\left\{[g^0],[g^2],[g^4]\right\}.
\label{eq:Com_Z10_Z2}
\end{equation}
and it stabilizes already at $L=1$, i.e., $\mathrm{Com}\left(\mathbb{Z}_{10}/\mathbb{Z}_{2}\right)^{\infty}=\mathrm{Com}\left(\mathbb{Z}_{10}/\mathbb{Z}_{2}\right)$. This implies that the all-loop exact groupified symmetry can already be identified at one loop, without analyzing higher-loop dressings. Let us now analyze how the elements are identified by dressing them with elements of $\mathrm{Com}\left(\mathbb{Z}_{10}/\mathbb{Z}_{2}\right)$, where only $[g^2]$ and $[g^4]$ are relevant.
By dressing with $[g^2]$, one finds
\begin{equation}
\begin{aligned}
[g^2]&\prec [g^2]\otimes[g^0],\quad 
[g^4]\prec [g^2]\otimes[g^2],\\
[g^3]&\prec [g^2]\otimes[g^1], \quad
[g^5]\prec [g^2]\otimes[g^3].
\end{aligned}
\label{eq:Z10_dressing_g2}
\end{equation}
Similarly, by dressing with $[g^4]$, one finds
\begin{equation}
[g^4]\prec [g^4]\otimes[g^0],\quad 
[g^5]\prec [g^4]\otimes[g^1].
\label{eq:Z10_dressing_g4}
\end{equation}
They identifies $[g^0]\sim[g^2]\sim[g^4]$ and $[g^1]\sim[g^3]\sim[g^5]$ at one-loop order, respectively.
Thus, while all elements are distinct at tree level, one-loop dressing identifies elements within each parity class, leaving only the even and odd classes distinguishable. It follows that at one-loop order the fusion algebra reduces to $\mathbb{Z}_2$, and this is the all-order exact group as the commutator set already stabilizes at $L=1$.

Notice that the above analysis relies only on the fusion rules, without any inspection of a Lagrangian. By contrast, the corresponding classical renormalizable Lagrangian contains more than $60$ interaction terms; although one could still identify the symmetry by direct inspection, the result would not emerge in such a clean and systematic way.

\bibliographystyle{elsarticle-num}
\bibliography{Noninvert_SR}

\end{document}